\documentclass[prl,aps,twocolumn,noshowpacs,noshowkeys,superscriptaddress
]{revtex4-1}
\usepackage{amsfonts}
\usepackage{amsmath}
\usepackage{amssymb}
\usepackage{graphicx}
\usepackage{textcomp}%
\newcommand{\dxg}{\delta x_{\rm G}}
\newcommand{\dxs}{\delta x_{\rm S}}
\newcommand{\Ws}{\Omega_{\rm S}}
\newcommand{\Wg}{\Omega_{\rm G}}
\newcommand{\Gs}{\Gamma_{\rm S}}
\newcommand{\Gg}{\Gamma_{\rm G}}

\begin{document}

\def\neel{Institut N\'{e}el, Universit\'{e} Grenoble Alpes - CNRS:UPR2940, 38042 Grenoble, France}
\author{Cornelia Schwarz}
\affiliation{\neel}
\author{Benjamin Pigeau}
\affiliation{\neel}
\author{Laure Mercier de L\'epinay}
\affiliation{\neel}
\author{Aur\'elien G. Kuhn}
\affiliation{\neel}
\author{Dipankar Kalita}
\affiliation{\neel}
\author{Nedjma Bendiab}
\affiliation{\neel}
\author{La\"{e}titia Marty}
\affiliation{\neel}
\author{Vincent Bouchiat}
\affiliation{\neel}
\author{Olivier Arcizet}
\affiliation{\neel}
\email{olivier.arcizet@neel.cnrs.fr}

\title{Deviation from the normal mode expansion\\ in a coupled graphene-nanomechanical system}

\begin{abstract}
{\bf We optomechanically measure the vibrations of a nanomechanical system made of a graphene membrane suspended on a silicon nitride nanoresonator. When probing the thermal noise of the coupled nanomechanical device, we observe a significant deviation from the normal mode expansion. It originates from the heterogeneous character of mechanical dissipation over the spatial extension of coupled eigenmodes, which violates one of the fundamental prerequisite for employing this commonly used description of the nanoresonators' thermal noise. We subsequently measure the local mechanical susceptibility and demonstrate that the fluctuation-dissipation theorem still holds and permits a proper evaluation of the thermal noise of the nanomechanical system. Since it naturally becomes delicate to ensure a good spatial homogeneity at the nanoscale, this approach is fundamental to correctly describe the thermal noise of nanomechanical systems which ultimately impact their sensing capacity.}
\end{abstract}

\maketitle

Nanomechanical oscillators are routinely used in fundamental and applied physics \cite{Cleland2003,Ekinci2005} as ultrasensitive force or mass sensors due to their increased sensitivity to their environment. The understanding of dissipation at the nanoscale is the key ingredient towards extreme sensitivity operation.
Among others, carbon-based nanoresonators and novel 2D materials \cite{CastellanosGomez2015} have revolutionized the field of nanomechanics \cite{Sazonova2004,Chen2009,Zande2010,Chaste2012,Moser2013,Weber2014, Singh2014,Tavernarakis2014, Cole2015, Reserbat-Plantey2015,Liu2015} by pushing the oscillator dimensions down to a single atomic layer.
The extreme sensitivities achieved are ultimately limited by the thermal noise of the nanoresonators, which underlines the importance of correctly understanding and describing their Brownian motion.
The thermal noise of a vibrating nanomechanical system is commonly described using the normal mode expansion, which assumes that each eigenmode is driven by an independent fluctuating Langevin force, presenting no correlation with other eigenmodes.  However, this intuitive description only holds when the mechanical dissipation is homogeneously distributed in the system \cite{Saulson1990, Pinard1999}.  Otherwise, inhomogeneous damping can create dissipative coupling between eigenmodes, leading to a violation of their assumed independence.
Such deviations which have been reported on macroscopic devices \cite{Yamamoto2001,Conti2003}, are expected to be extremely important in nanomechanical systems,  since it becomes increasingly difficult to ensure and even measure a good spatial homogeneity over the entire nanosystem as its size is decreased. However, no deviations from the normal mode expansion were observed at the nanoscale up to now, despite the large variety of nanoresonators investigated.\\
\begin{figure}[t!]
\begin{center}
\includegraphics[width=0.99\linewidth]{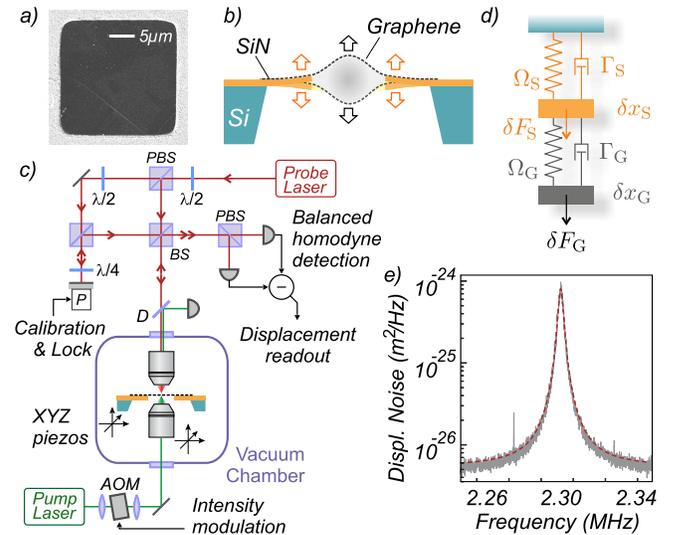}
\caption{
\textbf{ The experimental setup.} a) SEM of a $20\times20\rm \mu m^2$ suspended CVD grown graphene monolayer supported on a 300\,nm-thick SiN nanomembrane, as sketched in b). c) Experimental setup: a balanced homodyne detection measures the phase fluctuations of the probe laser field reflected by the sample and monitor its position fluctuations. A second counter-propagating pump laser beam can be intensity modulated to optomechanically drive the coupled nanoresonators. The experiment is performed  at pressures below $10^{-3}$\,mbar. d) Model describing the inertially coupled nanoresonators. e) Thermal noise of a graphene membrane, the sharp peaks on each side are weakly coupled SiN eigenmodes.}
\label{Fig1}%
\end{center}
\end{figure}
In this article, we report on the deviation from the normal mode expansion in the optomechanically measured thermal noise of a nanomechanical arrangement made of a suspended graphene monolayer coupled to a silicon nitride nanomembrane which supports the graphene resonator.  To fully explore the deviation from the normal mode expansion, we exploit the inertial coupling between both nanoresonators: under a temperature-controlled  tunable hybridization, the coupled eigenmodes become spatially delocalized on the two subsystems whose intrinsic mechanical damping rates differ by 2 orders of magnitude. In this situation the damping homogeneity is therefore altered which results in a pronounced deviation from the normal mode expansion that we report on and analyze. Then we prove that the fluctuation-dissipation theorem still holds by measuring the local mechanical susceptibility of the coupled nanomechanical system.\\
These considerations are essential to correctly describe nanomechanical systems affected by inhomogeneous damping and point out the importance of having access to the local mechanical susceptibility to correctly estimate the thermal noise of complex nanomechanical systems.\\

{\it The experiment--}
Our nanomechanical system is a fully suspended single layer graphene sheet deposited on a square window opened in a $\rm Si_3N_4$ nano-resonator, itself supported on a tapped silicon wafer (see Fig.\,1) which allows a dual optical access from both sides. It is obtained (see SI) by transfer in liquid phase of monolayer, poly-crystalline graphene grown by CVD on Cu \cite{Han2014,Reserbat-Plantey2014} and suspended over up to 25x25\,$\mu\rm m^2$ on a pre-etched stoichiometric $\rm Si_3N_4$ membrane which is 500-nm-thick and 100-$\rm\mu m$-wide.  A 633\,nm  probe laser is focussed on the graphene resonator with a high numerical aperture objective ($\approx400\,\rm nm$ optical waist). The weak reflected beam constitutes the signal arm  of a balanced homodyne detection \cite{Cohadon1999} (see Fig.\,1c and SI). This interferometer permits a shot-noise limited readout of the membrane's thermal noise with injected optical powers ranging from 1 to $100\,\rm \mu W$.  A fast piezo element driving the local oscillator mirror permits a robust calibration of the interferometer, insensitive in particular to spatial drifts or reflectivity variations due to non-homogeneous graphene properties (wrinkles or grain boundaries). Reflectivities in the $1-10\%$ range were measured on monolayers depending on the level of contaminants.  A typical calibrated displacement noise spectrum is shown in Fig.\,1. Its reproduction at varying optical powers permits to verify the absence of optical backaction (see SI).
The uncoupled graphene resonators present fundamental eigenmodes in the 1-10 MHz range, with quality factors from 10 to 500 in vacuum and effective masses ranging from $10^{-16}$ to $10^{-14}\,\rm kg$.
Operating with fully transmitting systems permits suppressing additional cavity effects \cite{Barton2012} which could complicate the noise thermometry. The spatial profile of graphene eigenmodes can be mapped by probing thermal noise spectra at varying positions on the graphene membrane, see SI. The slight elliptical structure and the frequency splitting observed on higher order modes reflects the presence of a residual 20 MPa stress along the diagonal direction \cite{Fartash1992,Seitner2014}, attributed to the graphene transfer process. Also visible on the thermal noise spectrum are sharp peaks corresponding to higher order eigenmodes of the $\rm Si_3N_4$ nanomembrane, whose fundamental mode oscillates around $100\,\rm kHz$.  They present larger quality factors (above 1000) but higher masses, on the order of $10^{-12}\,\rm kg$.
In the following we investigate the thermal noise of the coupled system.\\

{\it Hybridization of  graphene eigenmodes--} In order to tune the eigenfrequencies, we exploit the partial absorption of a second laser beam at 532\,nm focussed down to an optical waist of $\simeq300\,\rm nm$, spatially superimposed on the probe beam and injected from the backside of the sample. It generates a slight temperature increase which is almost not detectable in the Brownian temperature (see Fig.\,2e) but is sufficient to significantly thermally tune the graphene eigenfrequency. A clear hybridization between both graphene and $\rm Si_3N_4$ eigenmodes is shown in Fig.\,2b where a pronounced frequency anticrossing can be seen, as well as a modification of the mechanical damping rates. Such signatures are fingerprints of strong dual mode coupling \cite{Novotny2010}, which can also affect the force sensitivity \cite{Anetsberger2008,Jockel2011,Tsaturyan2014}.\\
The modelisation of our inertially coupled nanomechanical system is based on cascaded mechanical oscillators \cite{Saulson1990}, as sketched in Fig.\,1d. Their vibrations $\dxg, \dxs$  around the rest positions are coupled through $\ddot{\dxg} =-\Wg^2 \left(\dxg-\dxs\right)-\Gamma_{\rm G} \, (\dot\dxg- \dot\dxs) + {\delta F_{\rm G}}/{M_{\rm G}}$ and $
\ddot{\dxs} =-\Ws^2 \,\dxs-\Gamma_{\rm S} \, \dot\dxs+\mu\Wg^2 \left(\dxg-\dxs\right)   +\mu\Gamma_{\rm G} \, (\dot\dxg- \dot\dxs)+ {\delta F_{\rm S}}/{M_{\rm S}}$, where $\Omega_{\rm G,S}/2\pi$ (resp. $\Gamma_{\rm G,S}$ ) are the uncoupled frequencies (resp. damping rates).  $\delta F_{\rm G}$ is an external force applied on the graphene membrane, $M_G$ the graphene effective mass at the measurement location \cite{Pinard1999}, while $\mu\equiv M_G/M_S$ parameterizes the hybridization strength. Depending on the graphene and $\rm Si_3N_4$ membrane geometries which govern the vibration mode spectrum and their spatial profiles, anticrossings with varying strength can be observed (see SI). Intuitively, if graphene is positioned at a node of the membrane eigenmode, their hybridization will be reduced.
In the Fourier domain we have  $\left(\begin{smallmatrix}\delta x_{\rm G}\\ \delta x_{\rm S} \end{smallmatrix}\right)= {\boldsymbol \chi}[\Omega]\cdot \left(\begin{smallmatrix}\delta F_{\rm G}\\ \mu\, \delta F_{\rm S} \end{smallmatrix}\right)$, using $\delta x_i[\Omega]\equiv\int_\mathbb{R}{dt e^{i\Omega t} \delta x_i(t) }$.
The dynamical matrix $ {\boldsymbol \chi}[\Omega]^{-1}$ being
\begin{equation}
\left(
\begin{array}{cc}
\chi_{\rm G}^{-1}
&
M_{\rm G}\Omega^2-\chi_{\rm G}^{-1}
\\
\mu\left(M_{\rm G}\Omega^2-\chi_{\rm G}^{-1}\right)
&
\mu\left(\chi_{\rm S}^{-1}+\chi_{\rm G}^{-1}+ M_{\rm G}\Omega^2\right)\\
\end{array}
\right),
\label{eq.admit}
\end{equation}
where we used the uncoupled mechanical susceptibilities $\chi_{\rm G,S}\equiv M_{\rm G,S}^{-1}\left(\Omega_{\rm G,S}^2-\Omega^2-i\Omega\Gamma_{\rm G,S}\right) ^{-1}$.
Diagonalizing the restoring force matrix ${M_{\rm G}^{-1}}{\boldsymbol \chi}[0]^{-1}$ yields the new eigenfrequencies $\Omega_\pm/2\pi$ of the coupled system:
\begin{equation}
\Omega_\pm^2\equiv\frac{(1+\mu)\Wg^2+\Ws^2}{2}
\pm\frac{\sqrt{\left( \Ws^2-(1+\mu)\Wg^2\right)^2+4\mu\Wg^2\Ws^2}}{2}\label{eq.freq}.
\end{equation}
When $\mu\ll1$, the minimum relative frequency splitting amounts to $\sqrt{\mu}$, corresponding to a canonically defined coupling strength of $g=\Wg\sqrt{\mu}$  \cite{Novotny2010}. Depending on the sample geometry a large variety of coupling strengths can be observed, up to $200\,\rm kHz$, largely entering the so-called strong coupling regime ($g> \Gs,\Gg$). The experimentally measured coupled eigenfrequencies are shown in Fig.\,2c for increasing pump laser powers.  They can be well fitted using equation (\ref{eq.freq}) and a linear pump power dependence for the uncoupled graphene and $\rm Si_3N_4$ eigenfrequencies of $-284\,\rm Hz/\mu W$  and $-2\,\rm Hz/\mu W$  respectively. The latter corresponds to a maximum static heating of the $\rm Si_3N_4$ nanoresonator estimated at the level of $\simeq 1\,\rm K$ \cite{Larsen2011}. Using the experimentally measured heat diffusion coefficient of $5\times10^{-6}\,\rm m^2/s$, see SI, the thermal heat resistance of graphene was numerically estimated at  $0.25\,\rm K$ per $\rm \mu W$ absorbed. The effective mechanical damping rates $\Gamma_\pm$ of the coupled modes can be roughly estimated using the FWHM of the thermal noise spectra, see Fig.\,2d, and used to extrapolate the uncoupled damping rates (see SI).\\
\begin{figure}[t!]
\begin{center}
\includegraphics[width=0.99\linewidth]{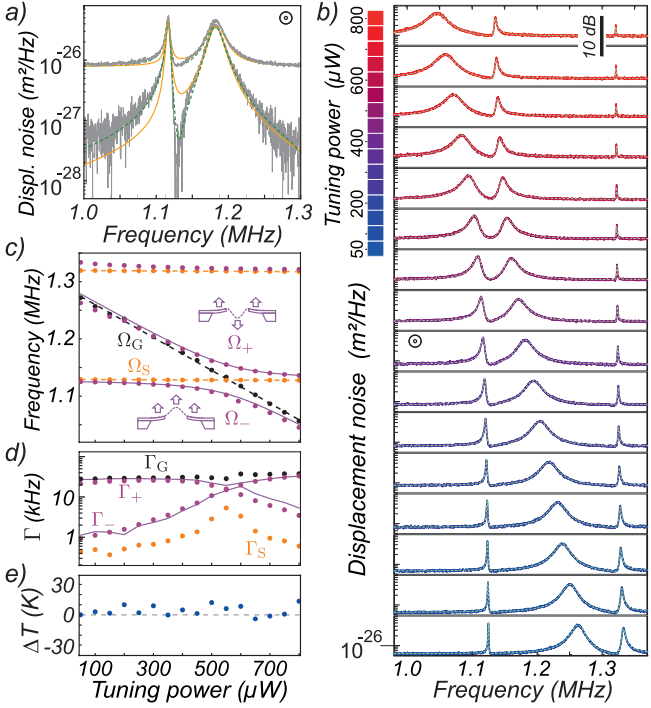}
\caption{
\textbf{ Thermal noise of the hybridized eigenmodes} a) Thermal noise of the coupled nanomechanical system when tuned to an anticrossing region by adjusting the pump intensity ($400\,\rm\mu W$). Lower traces are obtained after numerical background substraction. Solid lines are the best fits derived employing the normal mode expansion. Dashed green lines are fits using expression (\ref{eq.sxg}). b) Spectra measured through the anticrossing for increasing tuning laser powers. Dashed lines are fits using  (\ref{eq.sxg}) with fitting parameters $\Omega_{\rm S,G}, \Gamma_{\rm S,G}$ reported in c), d) using $\mu=0.002$. Purple disks represent the measured coupled eigenfrequencies $\Omega_\pm/2\pi$  and solid lines are deduced from equation (\ref{eq.freq}). d) Similar analysis for damping rates $\Gamma_\pm/2\pi$.  e) Relative Brownian temperature variation deduced from the fits.}
\label{Fig1}%
\end{center}
\end{figure}

{\it Violation of the normal mode expansion--}
Meanwhile, a striking feature can be seen in the displacement noise spectra shown in Fig.\,2: a characteristic peak asymmetry and a sharp noise minimum between both eigenmodes are clearly visible in the anticrossing region. These spectra cannot be fitted with two independent mechanical thermal noise spectra, see Fig.\,2a,  with a deviation larger than 10 dB observed in the vicinity of $\Ws$. Therefore the measured thermal noise cannot be described by two eigenmodes driven with independent Langevin forces, which reveals the violation of the normal mode expansion. This is a consequence of the spatial inhomogeneity of damping rates across the system: acoustic vibrations are more efficiently damped in graphene than in $\rm Si_3N_4$. When the eigenmodes become hybridized, their spatial profiles are delocalized over both systems, see Fig.\,2c insets, so that mechanical damping becomes inhomogeneous over the eigenmode spatial extension. Thus the spatial profile of the vibration pattern cannot be stationary anymore since it is non-homogeneously damped and cannot be preserved over time. As such, dissipation is now able to couple eigenmodes, which breaks the fundamental hypothesis required to apply the normal mode expansion \cite{Saulson1990, Pinard1999}. When $\Wg=\Ws$, the thermal noise spectral density at the minimum noise frequency is measured at a level $\approx 2\Gg/\Gs$ times lower than the prediction of the normal mode expansion, see SI. The understanding of this deviation is critical to patch the normal mode expansion and work out an analytical description of the system fluctuations.\\

{\it Thermal noise of the hybridized nanomechanical system--}
To properly describe the nanosystem thermal noise, it is necessary to return to the original formulation of the fluctuation-dissipation theorem \cite{Kubo1966,Saulson1990}:
\begin{equation}
S_{\dxg}[\Omega]=\frac{2 k_B T}{\left|\Omega\right|}\left|{\rm Im} \chi_{\rm GG}[\Omega]\right|,\label{eq.sxg}
\end{equation}
which relates the measured displacement noise spectral density to the local mechanical susceptibility  $\chi_{\rm GG}$. The latter connects the optomechanically measured deformations of the graphene membrane $\dxg[\Omega]$ to the external force $\delta F_{\rm G}$ applied on the graphene membrane at the measurement point: $\dxg[\Omega]=\chi_{\rm GG}[\Omega] \, \delta F_{\rm G}$.
First we pursue the analysis based on the model employed above. Inverting equation (\ref{eq.admit}) we obtain:
\begin{equation}
\chi_{\rm GG}[\Omega]^{-1}= \chi_{\rm G}^{-1}- \frac{\left(\chi_{\rm G}^{-1}+M_{\rm G}\Omega^2\right)^2}{\chi_{\rm G}^{-1}+\chi_{\rm S}^{-1}+M_{\rm G}\Omega^2}.\label{eq.suscept}\\
\end{equation}
which permits deriving the expected thermal noise (see SI). Our experimental results can be well fitted with this model, see Fig.\,2a, 2b, using the fitting parameters which are reported in Fig.\,2c,\,2d and 2e. The magnitude of the coupling parameter $\mu=0.002$ is  also in agreement with the ratio of bare effective masses of both nanoresonators.  No significant variation in the fitted noise temperature could be detected, see Fig.\,2e,  which places an upper bound of $\simeq10\,\rm K$ on the  maximum temperature increase induced by the tuning laser. This observation is also consistent with the estimated thermal resistance given above and allows to neglect the role of temperature inhomogeneities in our modelisation. \\
\begin{figure*}[t!]
\begin{center}
\includegraphics[width=0.9\linewidth]{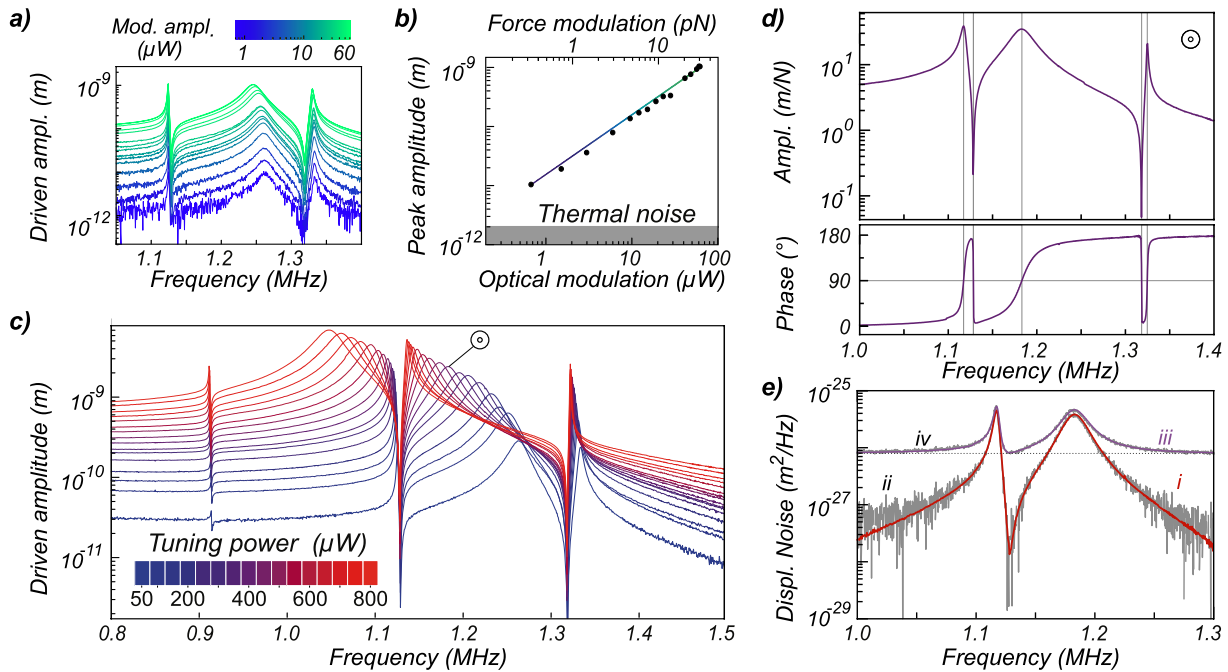}
\caption{ \textbf{ Optomechanical response of the hybridized nanomechanical system.} a). Optomechanical response obtained by modulating the pump intensity for increasing modulation depths $\delta P$ with a fixed average tuning power ($P_0= 60\,\rm \mu W$). b) Maximum driven displacement reported as a function of $\delta P$. The solid line has a slope of $17\,\rm \mu m/W$. c) Optomechanical responses obtained for increasing optical pump power $P_0$ ($30 \% $ modulation strength). d) Amplitude and phase of the mechanical susceptibility $\chi_{\rm GG}$ derived for $400\,\rm\mu W$ tuning power. The corresponding thermal noise spectrum expected using equation (\ref{eq.sxg}) is reported in e) (${\it i}$) and presents a very good agreement with the measured spectrum (${\it  ii}$). A $30\%$ correction was used here on the optical to force conversion factor determined in Fig.\, 3a, to account for a slight modification of the actuation efficiency between both measurements.}
\label{Fig3}%
\end{center}
\end{figure*}
{\it Validity of the fluctuation dissipation theorem in the coupled nanomechanical system--}
Verifying the validity of the fluctuation-dissipation theorem is essential in order to assess that the measured spectra correspond to the thermal noise of the system. Following the principles of linear response theory \cite{Kubo1966}, this requires measuring the local mechanical susceptibility $\chi_{\rm GG}$ of the coupled nanomechanical system. To do so we modulate the pump beam intensity by means of an acousto-optic modulator (AOM) and realize response measurements by sweeping the modulation frequency while recording the driven displacement.  Both laser spots are carefully superimposed on the graphene membrane to access to the local susceptibility; it is worth mentioning that this measurement cannot be realized with electrostatic gate or with piezo actuations since their spatial excitation profile is not localized on the measurement spot.  We first verify the linearity of the actuation, see Fig.\,3a by varying the optical modulation depth $\delta P$ over 2 orders of magnitude without modifying the mean pump power (60$\,\rm \mu W$) to ensure a stable graphene frequency, away from anticrossings. No deviation from linearity were observed in the driven oscillations up to a maximum amplitude of $1\,\rm nm$, a few times the monolayer thickness (0.3\,nm),  so that we perfectly sit in the linear actuation/measurement regime. A typical actuation efficiency of $17\,\rm pm/\mu W$ is measured, corresponding to an optical force of $540\,\rm fN/\mu W$. This is significantly larger than the radiation pressure force contribution of $0.3\,\rm fN/ \mu W$ for a $10\%$ absorption coefficient,  which confirms the dominant role of thermo-optical forces in the optical actuation of graphene \cite{Barton2012}. The backaction noise due to the intensity fluctuations of the shot noise limited laser beams can thus be evaluated at  the level of $\simeq0.1 \,\rm fm/\sqrt{Hz}$ for $P_0=100\,\rm \mu W$. This is largely negligible compared to the  measured thermal noise so that backaction cancelation \cite{Caniard2007} or classical noise squashing mechanisms \cite{Laurent2011} can be safely excluded to interpret our results.\\
Several response measurements were subsequently performed through the anticrossing in the same measurement conditions as in Fig.\,2a by progressively increasing the pump intensity, while maintaining a fixed modulation depth ($\delta P/P_0=30\%$). The response curves shown in Fig.\,3c permit, once combined with the optical to force conversion factor measured in Fig.\,3a in absence of hybridization, to determine the complex local mechanical susceptibility, $\chi_{\rm GG}[\Omega]$, as shown in Fig.\,3d. Its proper determination requires taking into account the weak residual contribution of the interferometer feedback loop in the measurement span, the transfer function of all photodetectors employed and the spectral response of the AOM. With this, the expected thermal noise can be properly estimated using equation (\ref{eq.sxg}) and compared to the measured thermal noise spectrum, as shown in Fig.\,3e. The excellent quantitative agreement found between both measurements all across the hybridization (see SI) demonstrates the validity of the fluctuation-dissipation theorem in our strongly coupled nanomechanical arrangement.\\
The hybridization dramatically modifies the graphene mechanical response and has an impact on the signal-to-noise ratio (SNR) observed in a force measurement. For a monochromatic force of amplitude $\delta F_{\rm G}$ applied in the center of the graphene membrane, the SNR can be expressed as $\frac{{\rm SNR}[\Omega]}{{\rm SNR}_{\rm G}}=\frac{|{\rm Im{\chi_{\rm G}^{-1}}}|}{|{\rm Im{\chi_{\rm GG}^{-1}}}|}$ where ${\rm SNR}_{\rm G}\equiv\delta F_{\rm G}^2/2M_{\rm G} \Gamma_{\rm G}k_B T$ represents the SNR of the uncoupled graphene alone. As verified experimentally and confirmed with the model, see SI, the SNR can be improved with respect to the uncoupled graphene resonator in narrow frequency bands in the vicinity of the $\rm Si_3N_4$ resonance. As already employed with macroscopic devices \cite{Conti2003}, this constitutes a strategy for achieving larger sensitivities in hybrid nano-sensors.\\

{\it Conclusions--} We have demonstrated the violation of the normal mode expansion in a multimode nanomechanical system and verified that the fluctuation-dissipation theorem well describes its thermal noise despite the large mass and damping asymmetries. This work underlines the importance of measuring the local mechanical susceptibility of a nanosystem to correctly understand its thermal noise. Since a good sample homogeneity cannot be maintained in extremely down-sized nanomechanical devices, we anticipate that these deviations will play an important role in the future of nanomechanical sensors. Our observations, realized on inertially coupled nanomechanical oscillators have a more general reach and are also valid when mechanical modes are externally coupled, such as by  optical or electrostatic force field gradients \cite{Shkarin2014,Gieseler2012, Faust2012a, Moser2013, Gloppe2014,Liu2015}. Such a fundamental approach could be used for developing new force detection protocols based on multimodal nanosystems.\\

\textit{Acknowledgements---} We warmly thank J.\,Jarreau, C.\,Hoarau, E.\,Eyraud,  D.\,Lepoittevin, B.\,Fernandez, J.P.\,Poizat, A.\,Gloppe and B.\,Besga for experimental and technical assistance. This project is supported by the ANR FOCUS, the ERC Starting Grant StG-2012-HQ-NOM, the E.U. Graphene Flagship and Lanef (CryOptics). C.S. acknowledges funding from the Nanoscience Foundation.

%

\end{document}